\RequirePackage{lineno} 

\documentclass[twocolumn,showpacs,preprintnumbers,amsmath,amssymb,nofootinbib]{revtex4}
\usepackage{graphicx}% Include figure files
\usepackage{dcolumn}% Align table columns on decimal point
\usepackage{bm}% bold math
 
\usepackage{epstopdf}

\def\bea{\begin{eqnarray}}
\def\eea{\end{eqnarray}}

\def\pp{\mbox{$p$-$p$}}

\def\pa{\mbox{$p$-A}}
\def\auau{\mbox{Au-Au}}

\def\pbpb{\mbox{Pb-Pb}}
\def\ppb{\mbox{$p$-Pb}}
\def\aa{\mbox{A-A}}
\def\nn{\mbox{N-N}}

\def\ppbar{\mbox{$p$-$\bar p$}}

\def\pt{$p_t$}
\def\yt{$y_t$}
\def\mpt{$\langle p_t \rangle$}
\def\nch{$n_{ch}$}

\begin{document} 

\setpagewiselinenumbers
\modulolinenumbers[5]
%\linenumbers

\preprint{version 0.3}

\title{
Ensemble mean $\bf p_t$ vs charged-hadron multiplicities in high energy nuclear collisions
}

\author{Thomas A.\ Trainor}\affiliation{CENPA 354290, University of Washington, Seattle, WA 98195}

%%%%%%%%%%%%%%%%%%%%%%%%%%%%%%%

\date{\today}

\begin{abstract}
Measurements of event-ensemble mean transverse momentum $\langle p_t \rangle$ vs charged-hadron multiplicity $n_{ch}$ for $p_t$ spectra from 5 TeV p-Pb and 2.76 TeV Pb-Pb collisions and from p-p collisions for several energies have been reported recently. While in all cases $\langle p_t \rangle$ increases monotonically with \nch\ the rate of increase is very different from system to system. Comparisons with several theory Monte Carlos  reveal substantial disagreements and lead to considerable uncertainty on how to interpret the \mpt\ data. In the present study we develop a two-component (soft+hard) model (TCM) of \pt\ production in high energy nuclear collisions and apply it to the \mpt\ data. The soft component is assumed to be a universal feature of high energy collisions independent of A-B system or energy. The hard-component model is based on the observation that dijet production in \pp\ collisions does not satisfy the eikonal approximation but does so in \aa\ collisions. Hard-component properties are determined from previous measurements of hadron spectrum hard components,  jet spectra and fragmentation functions. The TCM describes the \pp\ and \pbpb\ \mpt\ data accurately, within data uncertainties, and the \ppb\ data appear to transition smoothly from \pp\ to \aa\ \nch\ trends.
\end{abstract}


\maketitle

%%%%%%%%%
 \section{Introduction}
 
Proposed collision mechanisms that may determine hadron production in high energy nuclear collisions  are strongly debated.  Candidate mechanisms range from projectile nucleon dissociation and parton fragmentation to dijets~\cite{kn,hijing,ppprd,fragevo} to color reconnection of multiple parton interactions~\cite{pythia}, strong rescattering of partons and hadrons in a dense medium~\cite{ampt}, hydrodynamic flows~\cite{hydro,bw,epos} or a colored-glass condensate~\cite{kn,cgc}, possibly including glasma flux tubes~\cite{glasma}. Recent high energy data from the LHC and the emergence of apparently novel effects in \pa\ or d-A collisions have further stimulated debate.
 
 It is safe to assume that there is a direct connection between underlying collision mechanisms and the structure of the hadronic final state in yields, spectra, correlations, and especially jets as hadron correlations. However, how to characterize the final-state structure statistically and how to interpret it in terms of physical mechanisms remains an open question. All analysis methods are susceptible to bias, the information in the data may not be fully exploited, and physical mechanisms are conventionally represented by Monte Carlo (MC) simulations based on  complex assumptions that may be questioned.
 
 In the present study we emphasize the first moment of the transverse momentum spectrum or event-ensemble mean \pt\ denoted by $\langle p_t \rangle$, its variation with collision system A-B, collision energy $\sqrt{s_{NN}}$ and charged-hadron multiplicity \nch. Some previous studies  include \mpt\ vs \nch\ measurements from \pp\ collisions at the Sp\=pS~\cite{ua1mpt} and RHIC~\cite{ppprd}, measurements of event-wise \mpt\ fluctuations~\cite{ptfluct} and scale variation of those fluctuations in \auau\ collisions~\cite{ptscale,ptedep}, and variations of ensemble \mpt\ with centrality in  the \auau\ system~\cite{hardspec}. Here we consider recent \mpt\ vs \nch\ data from the LHC for \pp, \ppb\ and \pbpb\ collisions~\cite{alicempt}.
 
 There are striking differences between \mpt\ vs \nch\ trends in \pp\ and \aa\ collisions. \mpt\ tends to increase rapidly with \nch\ in \pp\ collisions but much more slowly in \aa\ collisions. The \pa\ trend is intermediate. The trend of \mpt\ increasing with centrality in \aa\ collisions is conventionally interpreted in terms of radial flow~\cite{radialflow}. That interpretation suggests that the \mpt\ trend in \pp\ collisions could also be associated with radial flow, with possibly larger magnitude. However, by the same argument if an alternative jet-related mechanism dominates \mpt\ in \pp\ collisions that same mechanism could prevail in \aa\ collisions, misinterpreted there as radial flow~\cite{nohydro}.
 
 We consider \mpt\ vs \nch\ data for several energies from 200 GeV to 7 TeV and from \pp\ or \ppbar\ collisions, \pbpb\ collisions and \ppb\ collisions. We apply the two-component (soft+hard) model (TCM) of hadron production to \mpt\ data systematics. We establish that jet production in \pp\ collisions has a simple dependence on the multiplicity soft component. We construct a TCM for \mpt\ in \pp\ and \aa\ collisions  and compare  the TCM to data. We conclude that in all systems the increase in \mpt\ with \nch\ or \aa\ centrality is entirely due to jet production. Details of the \mpt\ systematics in \pp\ collisions follow the properties of independently-measured jet spectra and dijet production. And \mpt\ systematics in \aa\ collisions reflect previously-observed modification of jet structure in \pt\ spectra for more-central collisions. The \ppb\ trend is simply explained as transitioning from \pp\ to \aa\ behavior as the number of nucleon participants in collisions becomes significantly greater than 2.
 
%%%%%%%%%%%%%%%%
This article is arranged as follows:
Sec.~\ref{alice} introduces recent LHC \mpt\ measurements with some conventional and alternative interpretations.
Sec.~\ref{dijetprod} reviews the systematics of dijet production in 200 GeV \pp\ collisions.
Sec.~\ref{twocomp} describes two-component models for \mpt\ data from \pp\ and \aa\ collisions.
Sec.~\ref{ppdat} reviews \mpt\ data from \pp\ collisions for several collision energies compared to the \pp\ TCM.
Sec.~\ref{aadat}  reviews \mpt\ data from 2.76 TeV \pbpb\ collisions compared to the \aa\ TCM.
Sec.~\ref{padat} compares \mpt\ data from 5 TeV \ppb\ collisions with \pp\ and \pbpb\ data and corresponding TCMs.
Secs.~\ref{disc} and~\ref{summ} present Discussion and Summary.

%%%%%%%%%%%%%%%
\section{Recent LHC $\bf \langle p_t \rangle$ measurements} \label{alice}

Reference~\cite{alicempt} reports measurements of \mpt\ vs \nch\ for several collision systems at LHC energies, including \pp\ at 0.9, 2.76 and 7 TeV, \ppb\ at 5 TeV and \pbpb\ at 2.76 TeV. Whereas \mpt\ is calculated within a reduced \pt\ acceptance \nch\ is extrapolated to \pt\ = 0. It is observed that \mpt\ increases with \nch\ much more rapidly for \pp\ collisions  than for \pbpb\ collisions and that \ppb\ data are intermediate, following the \pp\ trend for smaller \nch\ and the form (but not magnitude) of the \pbpb\ trend for larger \nch. Reference is made initially to similar previous measurements~\cite{ua1mpt,ppprd} but no details are presented.  It is anticipated that the \mpt\ vs \nch\ trends will shed light on underlying collision and hadron production mechanisms.
In  this section we review some conventional interpretations presented in Ref.~\cite{alicempt} and possible alternatives relating to the TCM for high energy nuclear collisions.

\subsection{Conventional interpretations}

The \mpt\ vs \nch\ trend for 7 TeV \pp\ collisions is compared to the PYTHIA Monte Carlo~\cite{pythia}. Default PYTHIA 8 tune 4C (particular model-parameter selection) strongly disagrees with the \pp\ data and seems more compatible in form with the \pbpb\ results. However, when a so-called color reconnection (CR) mechanism is applied to multiple parton interactions (MPI) PYTHIA appears to describe the \pp\ data. The CR/MPI mechanism is referred to as a collective effect.

It is stipulated that any \mpt\ increase with \aa\ centrality is conventionally attributed to radial flow of a dense medium, consistent with the so-called blast-wave model of \aa\ \pt\ spectra for which one model parameter is $\beta_t$, the radial flow velocity~\cite{bw}. There should then be a direct relation between \mpt\ and $\beta_t$ through the \aa\ \pt\ spectrum. Some MC models of \aa\ collisions include radial and elliptic flow mechanisms~\cite{ampt,epos}, but no theory model can describe the \pbpb\ \mpt\ data of Ref.~\cite{alicempt}, all models deviating by tens of percent. In either \pp\ or \pbpb\ cases comparisons with linear-superposition models (of MPI or \nn\ collisions respectively) are said to fail. The large difference in \mpt\ trends for given \nch\ leads to questions about the role of \nch\ in proposed collision mechanisms.

From comparisons among the  three collision systems the argument is presented that since collectivity (flow) is assumed for \aa\ collisions in the form of a thermalized flowing dense medium, and some degree of collectivity may be present in \pp\ collisions via the CR/MPI mechanism the intermediate behavior of the \ppb\ data suggests that some form of collectivity may emerge  there as well. Does radial flow play a role in \ppb\ collisions, in \pp\ collisions? If \mpt\ is a measure of radial flow is flow larger in \pp\ collisions? Other LHC results such as a claimed same-side ``ridge'' at large $\eta$ in high-multiplicity \pp~\cite{cmsridge} and \ppb~\cite{ppbridge} collisions suggest that flows may play a role at the LHC even in the smallest collision systems.
     
\subsection{Alternative interpretations}

Such deliberations omit a substantial amount of established information about jet production and manifestations thereof in the context of the two-component model of high energy nuclear collisions. Jet production is the signature manifestation of QCD in high energy collisions. Some aspects of jet production are predicable via perturbative QCD (pQCD), and some are accurately known via a broad range of measurements. Given the combination it is possible to make quantitative predictions for jet manifestations in hadron yields, spectra and correlations.

For instance, substantial jet fragment contributions to 200 GeV \pp~\cite{ppprd} and \auau~\cite{hardspec} collisions are described quantitatively by a combination of measured fragmentation functions~\cite{eeprd} and minimum-bias (MB) jet spectra~\cite{jetmodel,fragevo}. The same hadron spectrum features attributed to radial flow~\cite{radialflow} are fully consistent with QCD jets for all \auau\ centralities~\cite{nohydro}. Certain jet-related structures in \pp\ and \auau\ angular correlations are also quantitatively related to single-particle spectrum structure by a QCD jet model~\cite{porter2,anomalous,jetspec}.

The TCM provides an overarching framework for such comparisons. It assumes that the major contributions to yields, spectra and correlations consist of two components: (a) nucleon dissociation to soft hadrons and (b) scattered-parton fragmentation to correlated jets. Successful implementation of the TCM requires appropriate differential analysis methods and an accurate and comprehensive description of MB jet production. A TCM reference is defined as linear superposition of a fundamental process: small-$x$ parton-parton interactions within \pp\ collisions or \nn\ interactions within \aa\ collisions. Deviations from a TCM reference may then reveal novelty in the composite system. The TCM has few parameters and a simple algebra, in contrast to most MC models. Below we apply the TCM to recent LHC \mpt\ data.

 %%%%%%%%%
  \section{Dijet production in $\bf p$-$\bf p$ collisions} \label{dijetprod}
  
  To construct a TCM for \mpt\ trends we must understand the systematics of dijet production in \pp\ collisions~\cite{ppprd,porter2,pptheory,jetmodel}. The TCM includes soft and hard hadron production such that $n_{ch} = n_s + n_h$ within some angular acceptance $\Delta \eta$. It is assumed that $n_s$ represents projectile nucleon dissociation and is proportional to the number of small-$x$ partons released in  the \pp\ collision. $n_h$ then represents the number of hadron fragments from dijets and is proportional to the number of dijets in the acceptance. Within that context \pt\ spectra can be decomposed into fixed soft and hard components as described in Ref.~\cite{ppprd}.
 
\begin{figure}[h]
   \includegraphics[width=1.65in,height=1.6in]{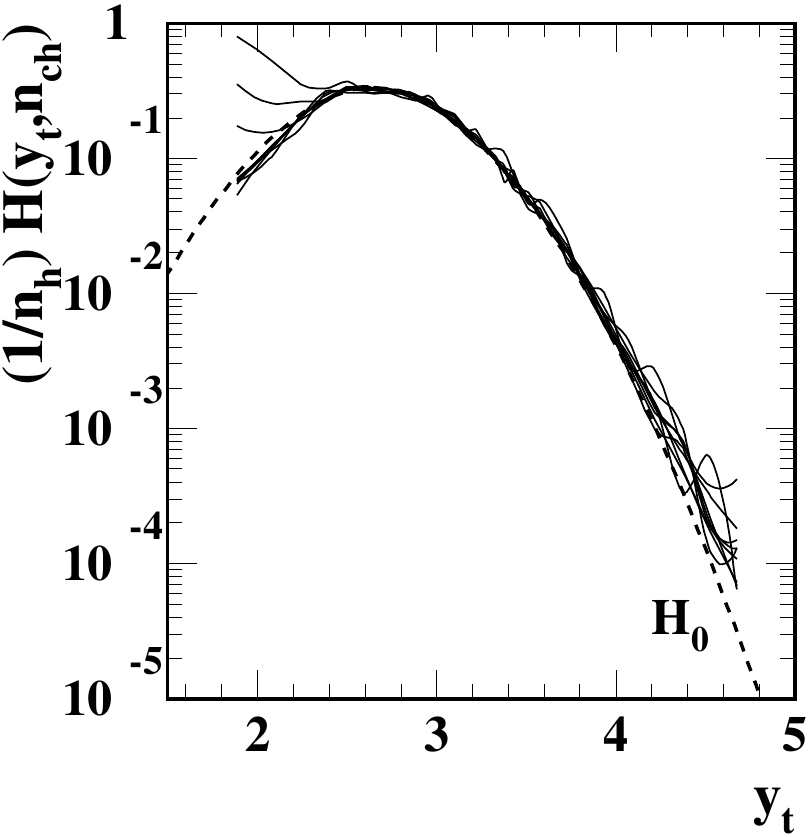}
    \includegraphics[width=1.65in,height=1.6in]{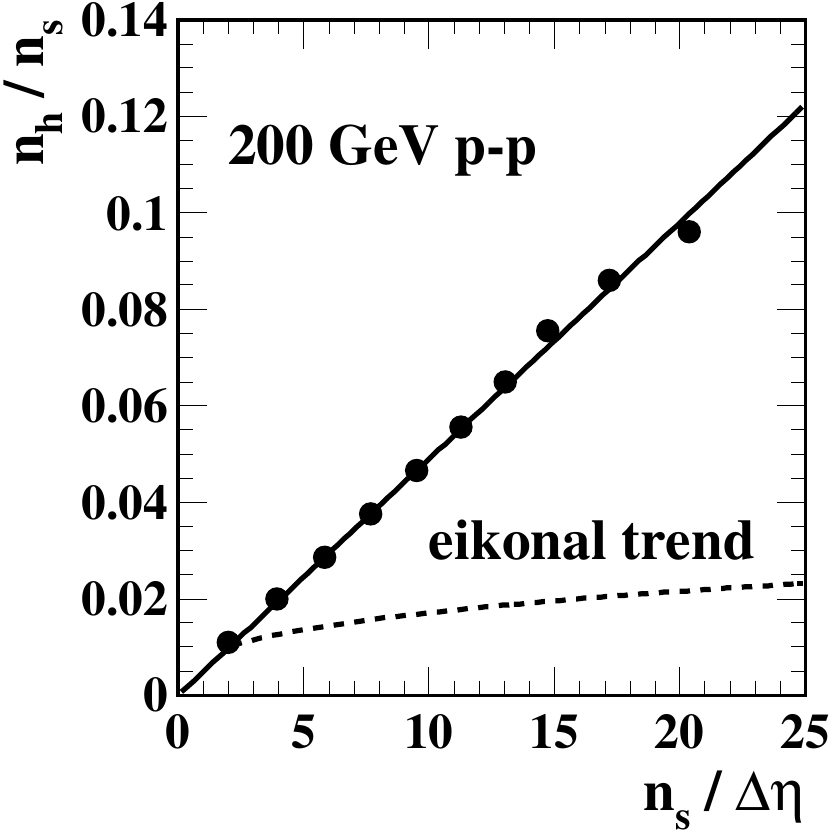}
 \caption{\label{ppprd}
 Left: Hard components $H(y_t,n_{ch})$ for transverse-rapidity \yt\ spectra from eleven multiplicity classes of 200 GeV NSD \pp\ collisions normalized by hard multiplicity $n_h$ within $\Delta \eta = 1$~\cite{ppprd}.  The dashed curve is unit-normal hard-component Gaussian reference $H_0$.
 Right: The ratio of hard to soft multiplicity (solid dots) plotted vs the soft multiplicity density. The trend indicates that dijets scale as $n_h = \alpha n_s^2$ (solid line) inconsistent with the trend $n_s^{4/3}$ (dashed curve) implied by the eikonal approximation. 
  } % ppcomm12bnew, 12cc
  \end{figure}

 Figure~\ref{ppprd} (left panel) shows \yt\ spectrum hard components $H(y_t,n_{ch})$ from ten multiplicity classes of 200 GeV non-single-diffractive (NSD) \pp\ collisions normalized by hard-component yield $n_h$ within acceptance $\Delta \eta = 1$~\cite{ppprd}. Transverse rapidity for unidentified hadrons is defined as $y_t = \ln[(m_t + p_t)/m_\pi]$. The full \yt\ spectra are described accurately by the sum of two fixed model functions $S_0(y_t)$ and $H_0(y_t)$ (soft and hard model components) with relative amplitudes $n_s$ (soft) and $n_h$ (hard)~\cite{ppprd}.  Deviations of data $H(y_t,n_{ch}) / n_h$ from model $H_0(y_t)$ below $y_t = 2.3$ (\pt\ = 0.7 GeV/c) for small \nch\ values are significant and play a role in the \mpt\ analysis below.
 
% $ ($\hat n_{ch} \approx 0.5 n_{ch}$ is the observed uncorrected multiplicity)
 
 %%%%%%%%%%
 %%%%%%%%%%%%
  
 Figure~\ref{ppprd} (right panel) shows the amplitudes of $H(y_t,n_{ch})/n_s$ vs soft multiplicity $n_s$. The line is $x = n_h / n_s = \alpha n_s$ consistent with $n_h = \alpha n_s^2$ for $\alpha \approx 0.006$ within acceptance $\Delta \eta = 1$.  Substantial evidence supports the interpretation that $n_s$ represents small-$x$ fragments from projectile-proton dissociation, and $n_h$ represents fragments from transverse-scattered-parton fragmentation~\cite{porter2,porter3}. That interpretation is consistent with QCD calculations derived from measured FFs and measured dijet cross sections~\cite{fragevo}. We then have a quantitative relation between hadron production via projectile dissociation and via scattered-parton fragmentation, with small-$x$ partons (mainly gluons) as the common element. 
 
% Based on the argument by analogy presented above we assume that soft hadron production follows a density distribution on longitudinal rapidity or pseudorapidity varying self-similarly with beam rapidity in the form $ \Delta y_b =  y_b - y_{b0}$, where $y_{b0}$ represents an energy cutoff scale $Q_0 \approx 10$ GeV discussed below. For the self-similar system we then expect $dn_s / d\eta \propto \Delta y_b$ and $n_{s,tot} \propto (\Delta y_b)^2$ in $4\pi$ consistent with $2 n_{ch,j} \propto (y_{max} - y_{min})^2$ for FFs.
 
% Given a spectrum hard component representing MB jets and the data in Fig.~\ref{ppprd} (right panel) we can write
% \bea
% dn_h/d\eta &=& f \epsilon(\Delta \eta) 2 \bar n_{ch,j} \approx 0.006 (dn_s/d\eta)^2 ,
% \eea
% where $f = dn_j/d\eta$ is the MB dijet density on $\eta$, $\epsilon(\Delta \eta) \in [0.5,1]$ is the fraction of a dijet that appears in $\Delta \eta$ and $2 \bar n_{ch,j}$ is the mean dijet fragment multiplicity within $4\pi$. Combined with the soft-component energy dependence above we have $f \propto (\Delta y_b)^2$ as the expected beam-energy dependence for MB dijet production at mid-rapidity.
 
 If $n_s$ is a proxy for participant small-x partons and MB dijet production scales accurately as $n_s^2$ we can conclude  that the number of binary parton-parton collisions is $N_{bin} \propto N_{part}^2$, where $N_{part}\sim n_s$ represents the number of participant small-$x$ partons in a \pp\ collision. The quadratic relation implies that any combination of participant partons can result in a large-angle dijet, {\em inconsistent with the eikonal approximation} where we expect $N_{bin} \propto N_{part}^{4/3}$ or equivalently $n_h/n_s \propto n_s^{1/3}$ as shown in Fig.~\ref{ppprd} (right panel, dashed curve).
 
 %%%%%%%%%
  \section{Two-component model for $\bf \langle p_t \rangle$} \label{twocomp}
  
  The TCM for yields spectra and correlations is based on the assumption that hadron production near mid-rapidity proceeds via soft or hard mechanisms assumed to be linearly independent. The soft component is assumed to be universal, the same for all systems and collision energies. The hard component follows a non-eikonal trend for \pp\ collisions and an eikonal trend for \aa\ collisions with larger A. The trend for \pa\ collisions is not known a priori but may involve a smooth transition from \pp\ to \aa. Within that context any increase in \mpt\ with \pp\ \nch\ or \aa\ centrality is due to jet production.  Given those relations between hard and soft components we now define a TCM for \mpt\ vs \nch.
  
 \subsection{$\bf \langle p_t \rangle$ TCM for p-p collisions}
 
 To define ensemble \mpt\ we begin with the concept of a total \pt\ denoted by $P_t$ integrated within some angular acceptance. If \nch\ is the total charge integrated within the same acceptance then $\langle p_t \rangle = P_t / n_{ch}$. Just as $n_{ch} = n_s + n_h$ we assume $P_t = P_{t,s} + P_{t,h}$. We then have $P_t = n_s \langle p_t \rangle_s + n_h\langle p_t \rangle_h$ and can define a TCM for \pp\ \mpt\
 \bea \label{pptcm}
  \langle p_t \rangle_{pp}(n_s)
  &=& \frac{n_s \langle p_t \rangle_{s} + n_h \langle p_t \rangle_{h}}{n_s + n_h} \\ \nonumber
  &=& \frac{\langle p_t \rangle_{s} + x (n_s)\langle p_t \rangle_{h}}{1 + x(n_s)}  \eea
  where $x(n_s) = \alpha n_s$ with $\alpha$ depending on the $\eta$ acceptance Given small adjustment of $\alpha$ for the experimental context we can derive $n_s$ from $n_{ch}$ by   $n_s = (1/2\alpha) [\sqrt{1+4\alpha n_{ch}} - 1]$, since $n_{ch} = n_s + \alpha n_s^2$. The two \mpt\ components can be inferred from the spectrum model functions $S_0(y_t)$ and $H_0(y_t)$ or from \mpt\ vs \nch\ data.
  
  The above relation assumes that the entire \pt\ spectrum is integrated to obtain \nch\ and $P_t$. If the spectrum is cut off at some small value $p_{t,cut}$ then
    \bea \label{pptcmp}
   \langle p_t \rangle'_{pp}(n_s,\sqrt{s}) &=& \frac{n'_s \langle p_t \rangle'_{s} + n_h \langle p_t \rangle_{h}(\sqrt{s})}{n_s' + n_h}
   \\ \nonumber
   &\approx&\frac{ \langle p_t \rangle_{s} + x(n_s) \langle p_t \rangle_{h}(\sqrt{s})}{n_s'/n_s + x(n_s)}
   \eea
where we assume no loss to the hard components of \nch\ and $P_t$. As we demonstrate in the next section $n'_s \langle p_t \rangle'_{s} \approx n_s \langle p_t \rangle_{s}$. That is, $P_{t,s}$ is relatively insensitive to a \pt\ cutoff provided $p_{t,cut}$ is sufficiently small (typical for cases relevant to this study). In that case the only effect  of the \pt\ cut is the ratio $n'_s / n_s$ in the denominator. We also note explicit energy dependence as anticipated.

\subsection{$\bf \langle p_t \rangle$ TCM for \aa\ collisions}

The TCM for \aa\ collisions is based on the Glauber model in which the fractional cross section (centrality) $\sigma/\sigma_0$ is related to geometry parameters $N_{part}$ the number of projectile nucleon participants, $N_{bin}$ the number of binary \nn\ encounters and $\nu = 2N_{part} / N_{bin}$ the mean participant pathlength in number of \nn\ encounters. The correspondence with observable \nch\ can be established from the minimum-bias cross-section distribution on \nch. For the present study the correspondence between Ref.~\cite{alicempt} \nch\ and Glauber model parameters was determined as described in Sec.~\ref{aadat}.

For \aa\ collisions the TCM of Eq.~(\ref{pptcm}) or (\ref{pptcmp}) must be modified in three ways: (a) the multiplicity hard component increases with centrality as $n_h(\nu)$, (b) due to modified parton fragmentation to jets in more-central \aa\ collisions the spectrum hard-component shape changes (softens) with centrality leading to variation of \mpt$_h$ as $\langle p_t \rangle_h(\nu)$ and (c) an \nn\ ``first encounter' effect must be accommodated, with details presented in Sec.~\ref{nnfirst}.

The direct extension of \pp\ $n_{ch} = n_s + n_h$ to  \aa\ is the first line of Eq.~(\ref{aanch}) where the \nn\ soft and hard components are scaled up by the corresponding Glauber parameters. However, the observed trend for hadron production in \aa\ collisions corresponds to $n_h$ for the first \nn\ encounter being the same as that for \pp\ no matter what the \aa\ centrality. For $\nu - 1$ subsequent encounters $n_h$ then transitions to a value depending on \aa\ centrality. The consequence is the second line that accurately describes \nch\ trends for a variety of collision systems
  \bea \label{aanch}
  n_{ch} &=& n_s (N_{part}/2) + \tilde n_h(\nu) N_{bin} 
  \\ \nonumber
  \frac{2}{N_{part}}n_{ch} &=& n_{pp} [1 + x(\nu) (\nu - 1)].
  \eea
  In that line $n_{pp} = n_s + n_{h,pp}$ and $x(\nu) = n_h(\nu) / n_{pp}$. Note that $\tilde n_h(\nu)$ is an average over all $\nu$ \nn\ encounters whereas $n_h(\nu)$ or $x(\nu)$ applies only to the  $\nu - 1$ subsequent encounters.

For a self-consistent description the same argument should be applied to $\langle p_t \rangle_{h}(\nu)$ such that in the first \nn\ encounter the \pp\ value holds while thereafter the value may change. The \aa\ TCM for \mpt\ with \pt\ cut is then
  \bea \label{aatcm}
 \langle p_t \rangle'_{AA} &=&  \frac{n_s \langle p_t \rangle_{s} (N_{part}/2) + \tilde n_h(\nu)  \tilde{\langle p_t \rangle}_{h}(\nu)N_{bin}}{n_s'(N_{part}/2) + \tilde n_h(\nu)N_{bin}} \\ \nonumber
 &=&  \frac{ \langle p_t \rangle_{s} + x_{pp}\, \langle p_t \rangle_{h,pp} + x(\nu)\, \langle p_t \rangle_{h}(\nu) (\nu-1)}{n_s'/n_s + x(\nu) \,(\nu-1)}.
  \eea
The Glauber model of \aa\ collisions based on the eikonal approximation gives the  trend $N_{bin} \sim N_{part}^{4/3}$ or $\nu \sim N_{part}^{1/3}$ for participant nucleons. The equivalent for \pp\ collisions is inconsistent with the eikonal approximation, with $n_h \sim N_{bin} \sim N_{part}^{2}$ or $\nu \sim N_{part} \sim n_s$, where $N_{part}$ in  the latter case represents participant small-$x$ partons.
 Equations~(\ref{pptcmp}) and (\ref{aatcm}) are employed below to analyze and interpret LHC \mpt\ data from Ref.~\cite{alicempt}.

 %%%%%%%%%%%%%%%
\section{LHC $\bf p$-$\bf p$ data} \label{ppdat}

We first consider \mpt\ vs \nch\  measurements for \pp\ collisions at three energies. We review the effects of applied \pt\ cuts based on a simple spectrum soft-component model that enables direct comparisons with full-acceptance results. We then apply the TCM to \mpt\ vs \nch\ data and confirm that the eikonal model cannot be applied to \pp\ collisions. We then extract the collision-energy dependence of the \mpt\ hard component. Comparison of that trend with the energy dependence of measured jet spectrum widths offers compelling evidence that \mpt\ variation is a manifestation of jet production in \pp\ collisions. 
  
\subsection{Effects of $\bf p_t$ acceptance cuts} \label{ptcut}

The \pp\ spectrum hard component $H(y_t)$ is negligible below $p_t \approx 0.35$ GeV/c (Fig.~\ref{ppprd}, left panel). A \pt\ cut imposed below that point affects only the soft component. We can calculate the consequences with  a soft-component model function. The unit-normal model function that describes spectrum soft components for both \pp\ and \auau\ collisions at 200 GeV is the L\'evy distribution
\bea
S_0(y_t) &=& \frac{20.4}{[1 + (m_t - m_h) / n T]^n},
\eea
where $m_h$ is the hadron mass (default is $m_\pi$), $T = 0.145$ GeV is the slope parameter and $n = 12.8$ is the L\'evy exponent, with $m_t = m_h \cosh(y_t)$ and $p_t = m_h \sinh(y_t)$.
  
\begin{figure}[h]
    \includegraphics[width=1.65in,height=1.6in]{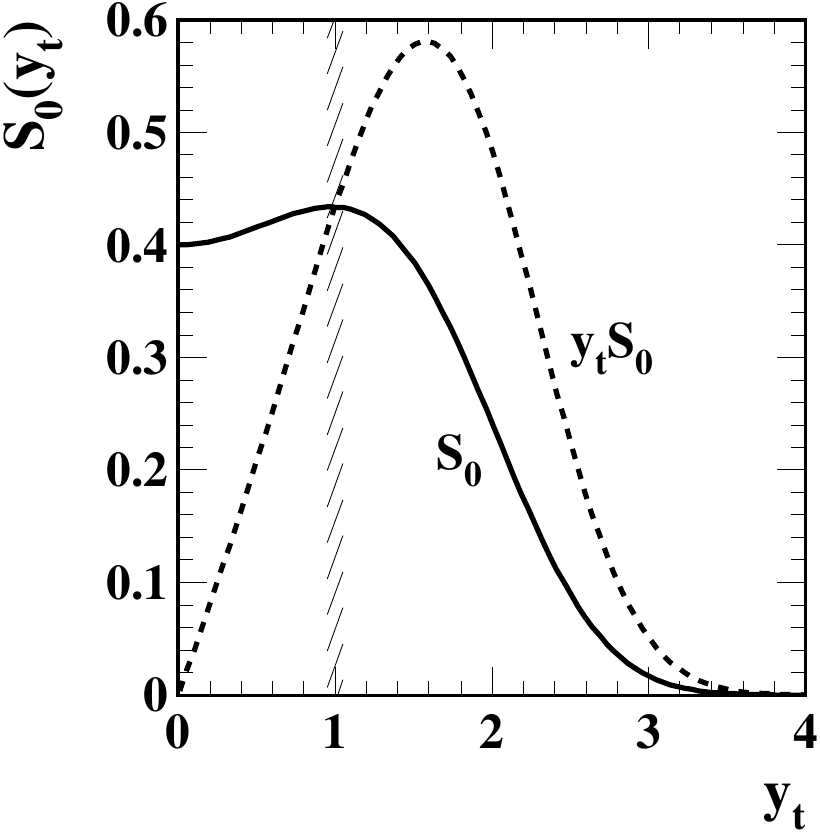}
   \includegraphics[width=1.65in,height=1.6in]{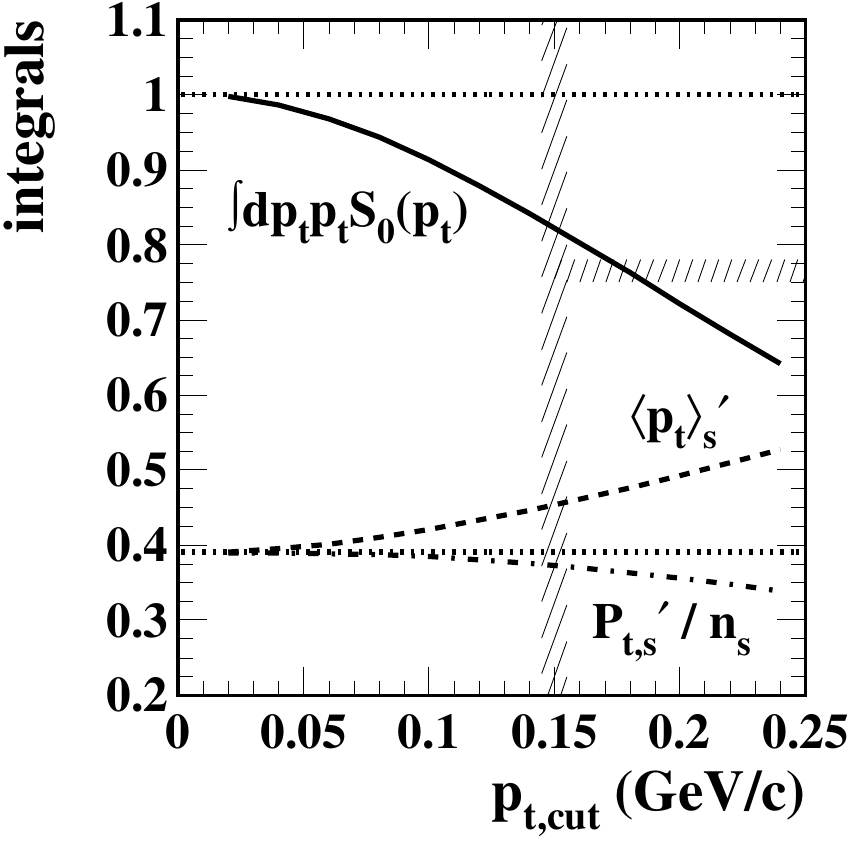}
  \caption{\label{ffs}
  Left: The \pp\ spectrum soft-component model $S_0(y_t) = (1/n_s) dn_s /y_t dy_t$ (solid curve) and product $y_t S_0(y_t)$ (dashed curve). The nominal \pt\ spectrum cutoff for Ref.~\cite{alicempt} at 0.15 GeV/c is denoted by the hatched band.
   Right: The integral of $p_t S_0(p_t)$ above $p_{t,cut}$ vs the cut value (solid curve). The nominal cut position is denoted by the vertical hatched band. The horizontal hatched band indicates the range of $n'_s / n_s$ ratios inferred from \mpt\ data. The effect of the cut on \mpt$_s$ is denoted by $\langle p_t \rangle'_s$ (dashed curve). The dash-dotted curve labeled $P'_{t,s} = n'_s \langle p_t \rangle'_s$ indicates that the $P_{t,s}$ product is much less sensitive to the \pt\ cut for smaller cut values.
     } %alice1c, 1b
   \end{figure}

 Figure~\ref{ffs} (left panel) shows unit-normal soft component $S_0(y_t)$ vs \yt\ (solid curve). Extrapolation to zero \pt\ is especially simple on transverse rapidity. For orientation \yt\ = 1, 2, 2.67, 3.35 and 4.05 correspond to \pt\ = 0.16, 0.5, 1, 2 and 4 GeV/c. The dashed curve shows $y_t S_0(y_t)$. The vertical hatched band indicates the nominal \pt\ acceptance cut for the Ref.~\cite{alicempt} data. About 18\% of the $y_t S_0(y_t)$ integral lies to the left of the band. However, the small-\nch\ intercepts of \mpt\ trends from Ref.~\cite{alicempt} indicate that the effective \pt\ cut may be somewhat higher.
 
   %%%%%%%%%%
  %%%%%%%%%%%%
 
  Figure~\ref{ffs} (right panel) shows the integral $\int_{p_{t,cut}}^\infty dp_t p_t S_0(p_t)$ (solid curve). The value is just the ratio $n'_s / n_s$ of accepted to true soft-component multiplicities. The vertical hatched band marks the nominal acceptance cut for the analysis in Ref.~\cite{alicempt}. The dashed curve shows the strong increase of $\langle p_t \rangle'_s$ with increasing \pt\ cut, whereas the dash-dotted curve shows the corresponding decrease in the total \pt\ soft component $P'_{t,s}$ (divided by the nominal NSD value of soft component $n_s$). The product $P_s = n_s \langle p_t \rangle_{s}$ is relatively insensitive to the \pt\ cut, and in what follows we assume that it does not change from the full-acceptance value.

 \subsection{p-p $\bf \langle p_t \rangle$ data}

\begin{figure}[h]
 \includegraphics[width=1.65in,height=1.6in]{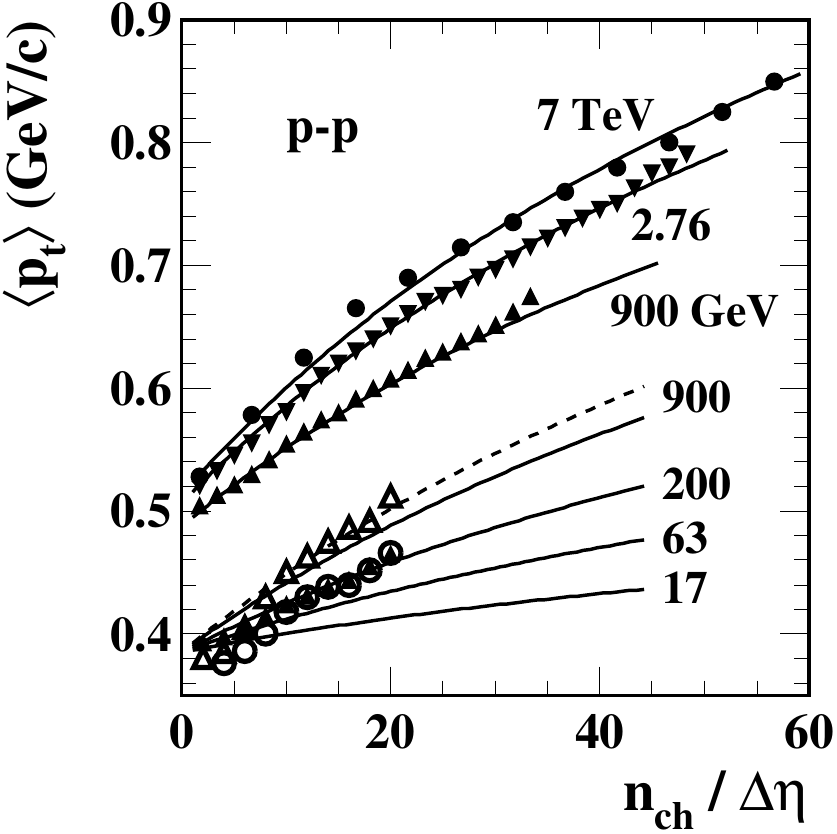}
 \includegraphics[width=1.65in,height=1.6in]{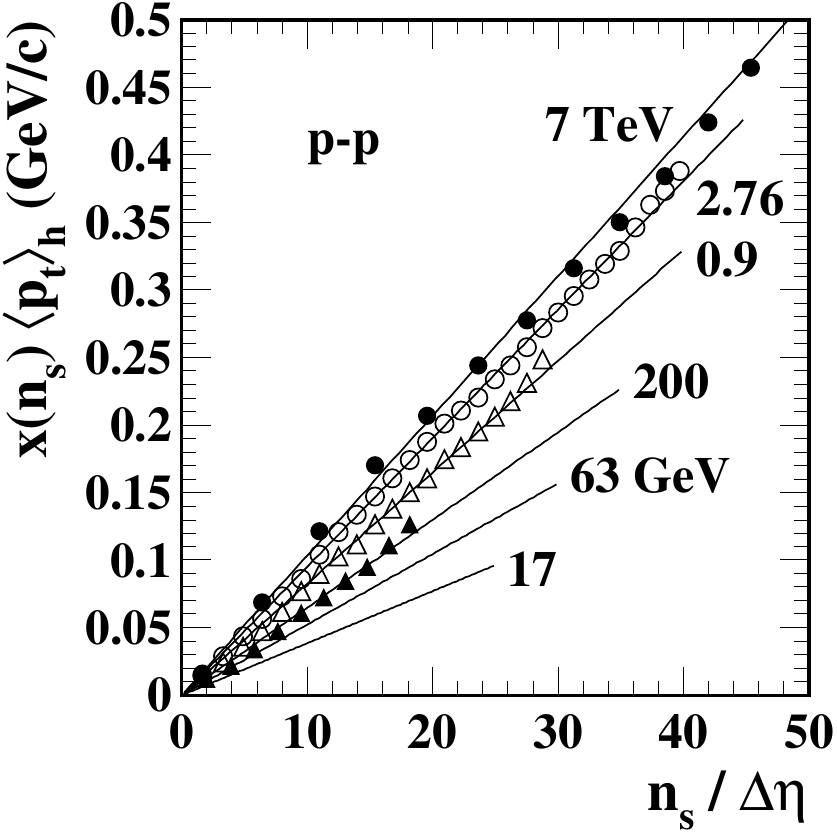}
\caption{\label{mptdat}
Left: \mpt\ vs \nch\ for several collision energies. The upper group of points is from Ref.~\cite{alicempt}. The lower 900 GeV data from UA1 derived from a ``power-law'' spectrum model~\cite{ua1mpt} fall substantially above the TCM for that energy (solid curve) but are consistent with the TCM form with amplitude adjusted (dashed curve).
Right: Data from the left panel multiplied by factor ${n'_{ch}} / n_s$ that removes the jet contribution and the effect of the \pt\ cut on the soft component from the denominator of \mpt. The universal soft component $\langle p_t \rangle_s$ is then subtracted according to Eq.~(\ref{mptrat}) isolating product $x(n_s) \langle p_t \rangle_h(\sqrt{s})$.
 }  %alice1a, 2bb
\end{figure}

Figure~\ref{mptdat} (left panel) shows LHC \mpt\ data from \pp\ collisions at 0.9, 2.76 and 7 TeV (upper points)~\cite{alicempt}. The $\langle p_t \rangle'$ values were calculated with a nominal $p_{t,cut} = 0.15$ GeV/c. Charge multiplicity \nch\ however was extrapolated to zero \pt. In Ref.~\cite{alicempt} an apparent change of slope near $n_{ch}/\Delta \eta = 16$ is noted and compared to similar claims from other experiments.  The curves are described below. Also included are lower-energy data from UA1 (open triangles, open circles~\cite{ua1mpt}) and STAR (solid points~\cite{ppprd}) for reference. The UA1 data for 900 GeV are  high compared to the overall energy trend. The UA1 analysis inferred \mpt\ values by fitting a ``power-law'' model function to \pt\ spectra. In Ref.~\cite{ppprd} possible biases arising from that method are discussed. As demonstrated below, the curvature of the \mpt\ vs \nch\ trends arises because a jet contribution is common to both numerator and denominator of \mpt. The trends should saturate at the hard-component value \mpt$_h(\sqrt{s})$ for large \nch.

%%%%%%%%%%
 %%%%%%%%%%%%
       
Figure~\ref{mptdat} (right panel) shows the quantity
\bea \label{mptrat}
\frac{n'_{ch}}{n_s} \langle p_t \rangle'(\sqrt{s}) -  \langle p_t \rangle_s &\approx & x(n_s)  \langle p_t \rangle_h(\sqrt{s})
\eea
 where the expression on the right follows from Eq.~(\ref{pptcmp}), $\langle p_t \rangle_s = 0.385$ GeV/c is assumed for all cases  and $x(n_s) = \alpha\, n_s / \Delta \eta$ with $\alpha = 0.0055$ for $\Delta \eta = 0.6$. The first term of  $n'_{ch}/n_s = n'_s/n_s + n_h / n_s$ is determined such that the various data sets have a common intercept point. The ratio $n'_s/n_s$ then has values within the horizontal hatched band in Fig.~\ref{ffs} (left panel) corresponding to $p_{t,cut} \approx 0.175$ GeV/c, slightly higher than the nominal cut value and suggesting some small uncorrected inefficiency near the acceptance boundary.  Variation of the line slopes is discussed in the next subsection. 
 
 \subsection{$\bf \langle p_t \rangle$ energy dependence and relation to MB jets}
 
\begin{figure}[h]
     \includegraphics[width=1.65in,height=1.6in]{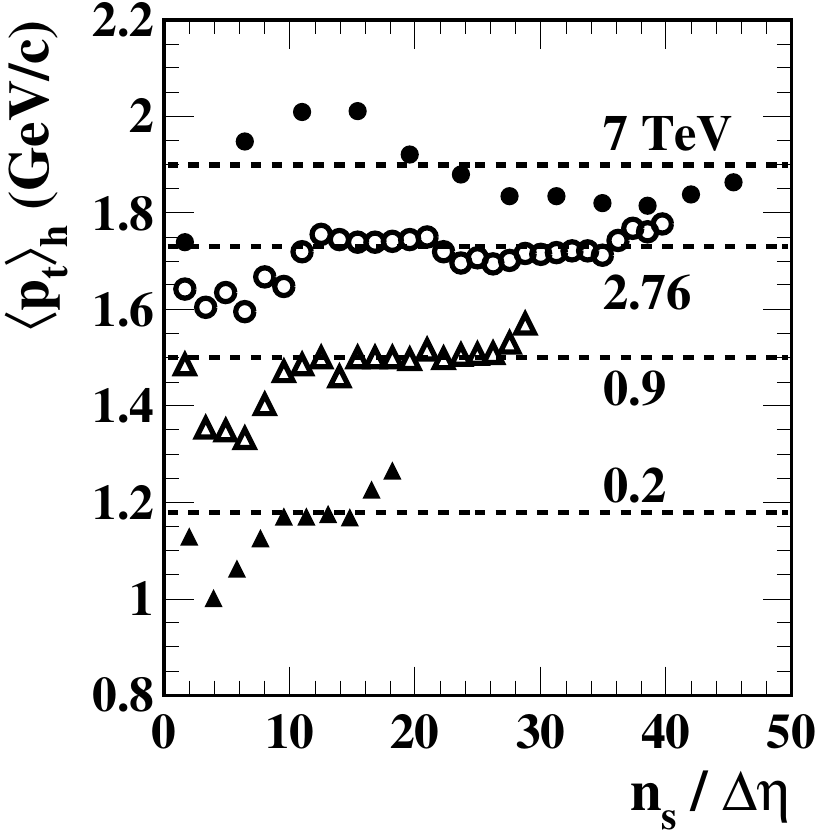}
     \includegraphics[width=1.65in,height=1.6in]{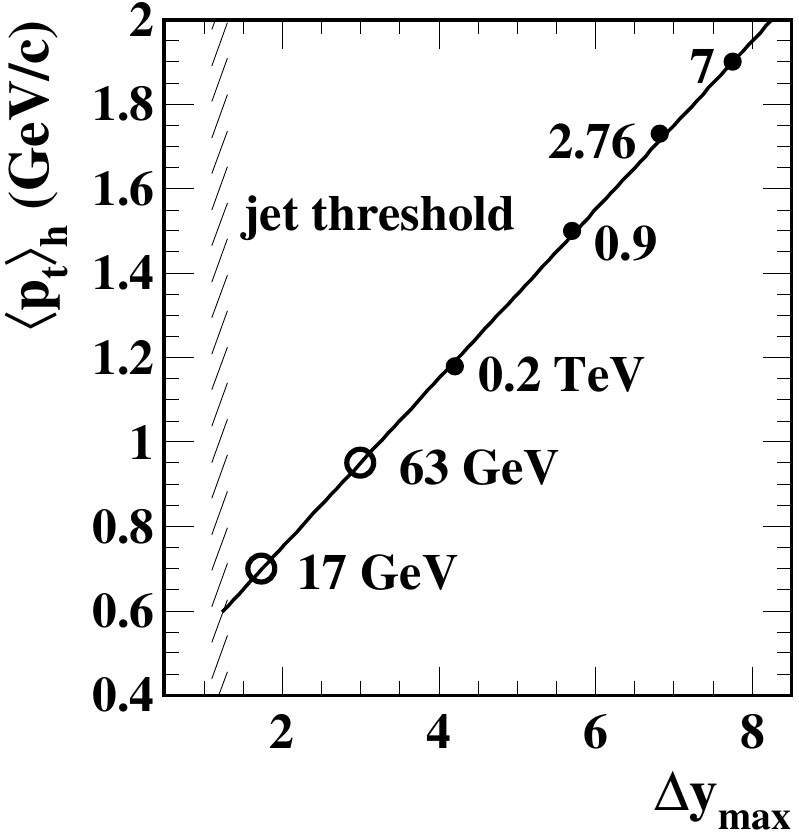}
   \caption{\label{endep}
   Left: \mpt\ hard components isolated according to Eq.~(\ref{mpthc}). The data fall within narrow horizontal bands reflecting the point-to-point data consistency with the TCM. The decreases for smaller $n_s$ reflect a real change in the hard component as seen in Fig.~\ref{ppprd} (left panel) for 200 GeV spectra, here shown to be common to a range of collision energies.
    Right: The $\langle p_t \rangle_h$ mean values from the left panel plotted vs parameter $\Delta y_{max} = \ln(\sqrt{s}/\text{3 GeV})$ that describes variation of the minimum-bias jet spectrum width with \pp\ collision energy~\cite{jetmodel}.
      }  %alice2c, 2d
    \end{figure}

 Figure~\ref{endep} (left panel) shows the \mpt\ data in the form
 \bea \label{mpthc}
\frac{1}{x(n_s)} \left(\frac{n'_{ch}}{n_s} \langle p_t \rangle'(\sqrt{s}) - \langle p_t \rangle_s\right) &=& \langle p_t \rangle_h(\sqrt{s})
 \eea
 for four energies, where $ \langle p_t \rangle_s$ has fixed value 0.385 GeV/c and $x(n_s)$ is defined above. Most of the $\langle p_t \rangle_h$ values fall in narrow horizontal bands, but the significant downturn for smaller multiplicities is a real change in the hard component for smaller \nch\ first observed in \pp\ spectrum hard components at 200 GeV, for instance in Fig.~\ref{ppprd} (left panel). The same effect apparently continues at least to 2.76 TeV. We find no evidence for a slope change in \mpt\ vs \nch\ near $n_{ch} / \Delta \eta = 16$ which would manifest in this figure as a significant step-wise decrease in $\langle p_t \rangle_h$ vs $n_s$.
 
   %%%%%%%%%%
   %%%%%%%%%%%%

 Figure~\ref{endep} (right panel) shows $\langle p_t \rangle_h(\sqrt{s})$ from the left panel (solid points) vs quantity $\Delta y_{max} = \ln(\sqrt{s} / \text{3 GeV})$ from Ref.~\cite{jetmodel}. In that study it is shown that jet spectrum widths scale with \pp\ collision energy as $\Delta y_{max}$. Thus we conclude from the right panel of this figure that $\langle p_t \rangle_h$ is linearly related to the minimum-bias jet spectrum width. That trend is also consistent with the results of Ref.~\cite{fragevo} where it is demonstrated that the spectrum hard component is predicted by folding an ensemble of fragmentation functions with a minimum-bias jet spectrum. In that case the hard-component width should scale linearly with the MB jet spectrum width, and $\langle p_t \rangle_h$ should have the linear correspondence to the jet spectrum width demonstrated above.  For \pp\ collisions the \mpt\ vs \nch\ systematics  compel a jet interpretation for the TCM hard component. The soft component remains consistent with a universal phenomenon independent of collision system or energy. The open symbols indicate predictions for lower energies. The hatched band represents the cutoff for dijet production near $\sqrt{s} = 10$ GeV determined from angular correlation analysis~\cite{ptedep,davidhq,anomalous}.

 %%%%%%%%%%%%%%%
 \section{LHC $\bf Pb$-$\bf Pb$ data} \label{aadat}

\begin{figure}[h]
  \includegraphics[width=1.65in,height=1.6in]{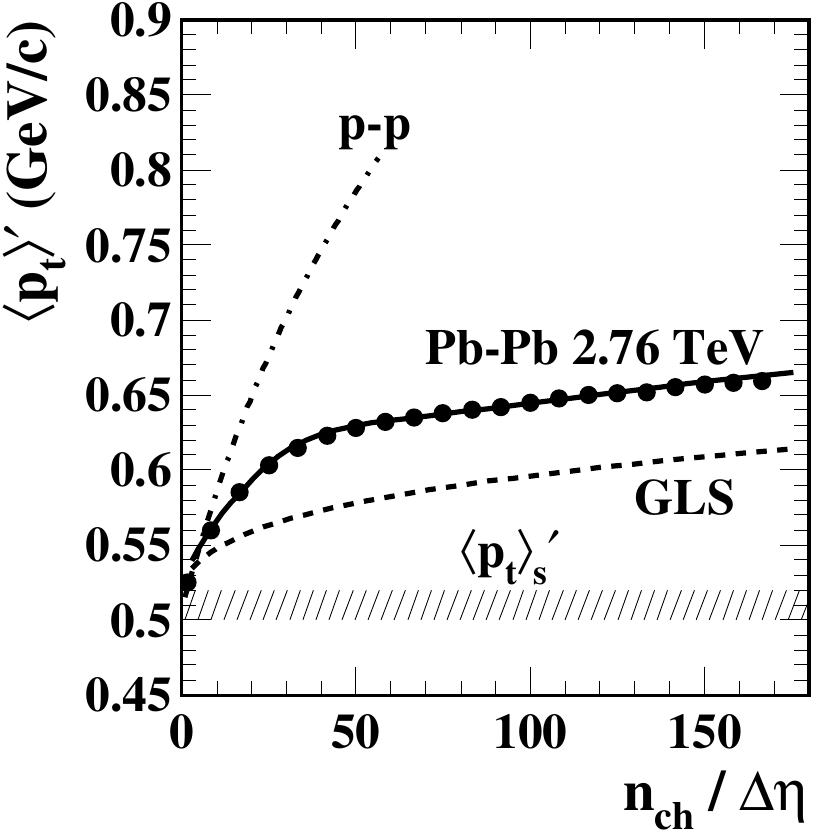}
  \includegraphics[width=1.65in,height=1.6in]{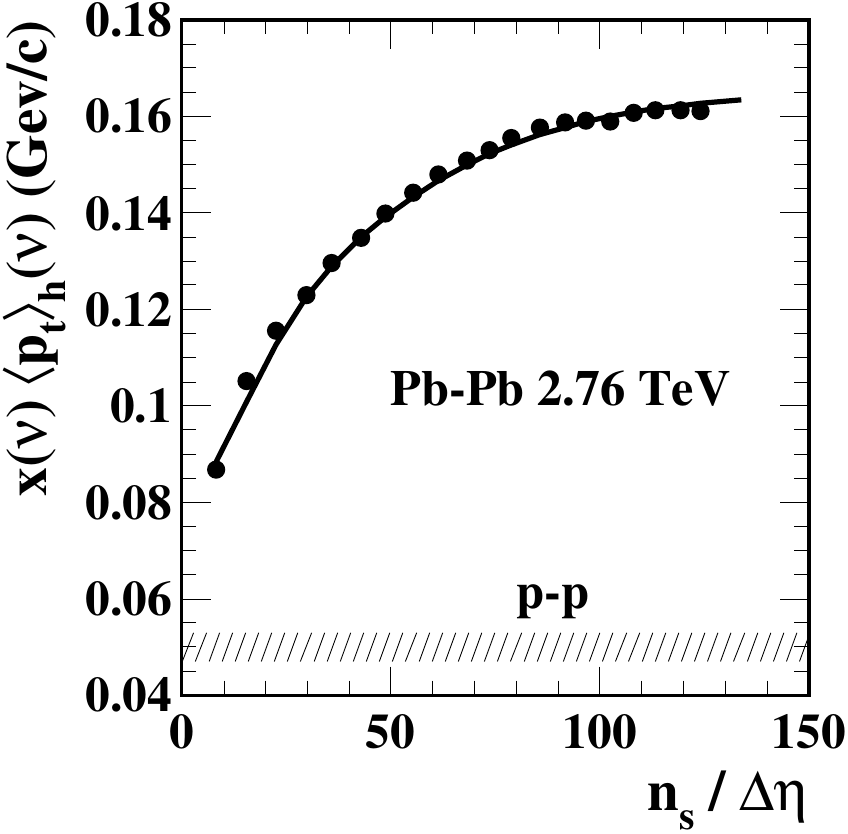}
  \caption{\label{mptaa}
  Left: \mpt$'$ data for 2.76 TeV \pbpb\ collisions from Ref.~\cite{alicempt} (points). The solid curve is the \pbpb\ TCM. The dashed curve is a Glauber linear superposition (GLS) reference assuming all \nn\ collisions are equivalent to \pp\ collisions and scale according to the Glauber model of \pbpb\ collisions following the eikonal approximation. $\langle p_{t}\rangle'_s = P_{t,s} / n'_{ch}$.
  Right: The product $x_h(\nu) \langle p_t \rangle_h(\nu) = P_{t,h}(\nu) / n_{pp}$ extracted from  data in the left panel according to Eq.~(\ref{prodd}) with $x_{pp} \langle p_t \rangle_{h,pp} \approx 0.05$ GeV/c. The solid curve is part of the TCM.
   }  %alice3ano, 3b
  \end{figure}

Figure~\ref{mptaa} (left panel) shows 2.76 TeV \pbpb\ \mpt\ data (points) from Ref.~\cite{alicempt} as samples from the full data complement. As noted in that study \mpt\ for \pbpb\ collisions increases much less quickly  than that for \pp\ collisions (the 2.76 TeV \pp\ trend is the dash-dotted curve). It is also noted that the \mpt\ increase in \aa\ collisions is conventionally attributed to radial flow corresponding to the blast-wave model applied to \pt\ spectra~\cite{radialflow}. The hatched band shows the \mpt\ soft component corresponding to $p_{t,cut} \approx 0.175$ GeV/c. The Glauber linear superposition (GLS) trend is Eq.~(\ref{aatcm}) with $x = 0.028$ and \mpt$_h = 1.75$ GeV/c fixed at their 2.76 TeV \pp\ values. The solid curve through data is discussed below.

  %%%%%%%%%%
  %%%%%%%%%%%%

Figure~\ref{mptaa} (right panel) shows the product $x(\nu) \langle p_t \rangle_h(\nu) = P_{t,h}(\nu) / n_{pp}$ (points) obtained from data in the left panel according to Eq.~(\ref{aatcm}) by  
 \bea \label{prodd}
x(\nu)\langle p_t \rangle_h(\nu) \hspace{-.02in} &=&\hspace{-.02in}\frac{   \frac{2}{N_{part}} \frac{n'_{ch}}{n_s} \langle p_t \rangle' -  \langle p_t \rangle_s - x_{pp}\langle p_t \rangle_{h,pp}}{\nu-1}.
 \eea
The hatched band shows the NSD \pp\ value $ x_{pp}\langle p_t \rangle_{h,pp} \approx 0.05$ GeV/c. The solid curve is discussed below. Those \pbpb\ results can be compared with Fig.~\ref{mptdat} (right panel) for \pp\ collisions. We next isolate the individual factors.
   
\begin{figure}[h]
 \includegraphics[width=1.65in,height=1.6in]{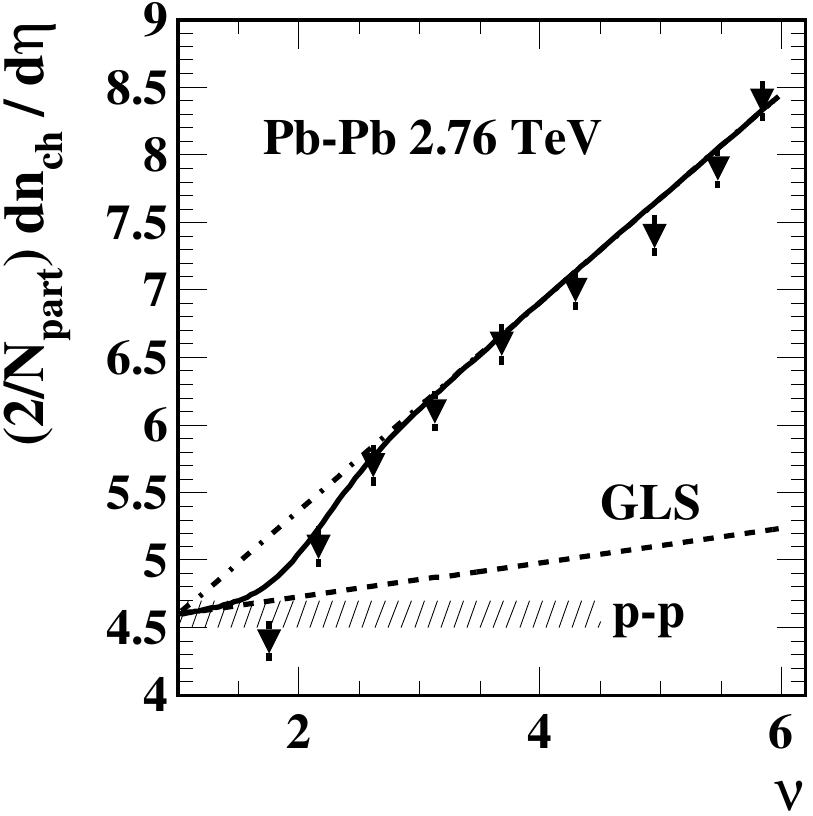}
 \includegraphics[width=1.65in,height=1.6in]{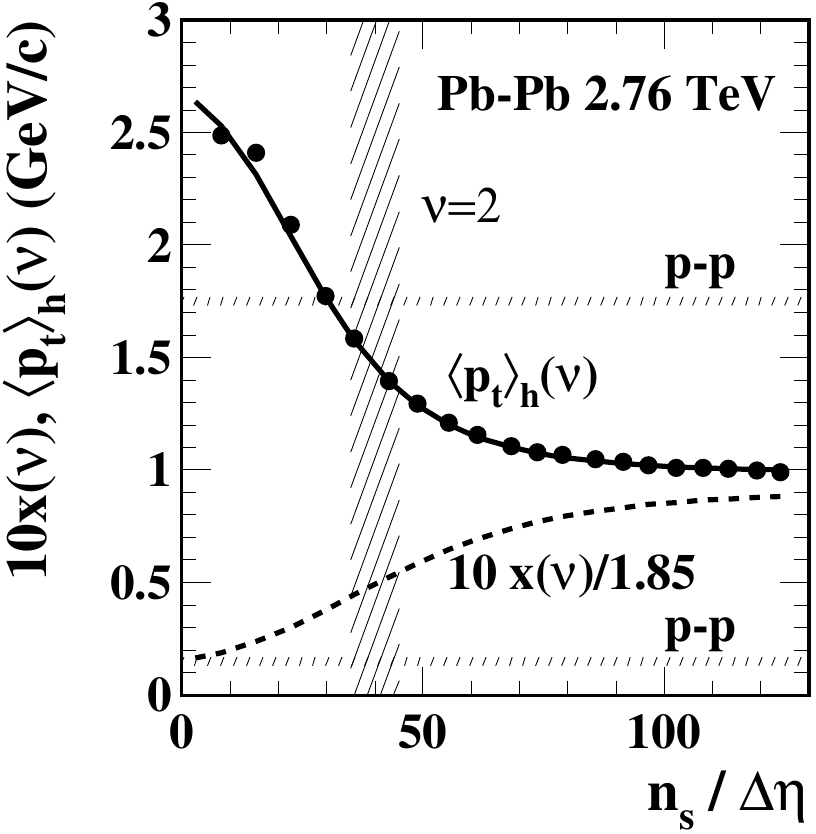}
 \caption{\label{alicench}
 Left: Hadron production data vs mean participant path length $\nu$ for 2.76 TeV \pbpb\ collisions from Ref.~\cite{alicench} (solid triangles). The lines and solid curve are the TCM for 200 GeV \auau\ collisions scaled up by factors 1.85 (soft component) and $1.85^2$(hard component) reflecting soft multiplicity $n_s$ scaling as $\ln(\sqrt{s} / \text{10 GeV})$, as noted in Refs.~\cite{tomphenix,jetmodel}.
 Right: The product data from Fig.~\ref{mptaa} (right panel) divided by ratio model $x(\nu)$ defined by Eq.~(\ref{xnual}) and shown as the dashed curve in this panel. The solid curve is the $\langle p_t \rangle_h(\nu)$ model defined by Eq.~(\ref{mptnu}). The combination is the basis for the \pbpb\ TCM.
  }  %alice3c, 3d
 \end{figure}

Figure~\ref{alicench} (left panel) shows 2.76 TeV \pbpb\ hadron production data from Ref.~\cite{alicench} (points) compared to the corresponding TCM in the general form 
  \bea \label{nchalice}
\frac{2}{N_{part}}  n_{ch} &=& n_{pp} [1 +  x(\nu) (\nu - 1)]
  \eea
  where for 200 GeV \auau\ $n_{pp} \approx 2.5$, $x(\nu) \in [0.015,0.095]$ and $x(1) = \alpha n_{s,NSD} = 0.015$. For 2.76 TeV the factor $1.85\approx  \ln(2760 / 10) / \ln(200/10)$ predicts the expected increase in $n_{s,NSD} \approx n_{pp} \rightarrow 4.6$ scaling with small-$x$ partons as described in Ref.~\cite{jetmodel}. The same factor is applied to $x(\nu)$ per Sec.~\ref{dijetprod} . The functional form of $x(\nu)$ at 2.76 TeV is very similar to that at 200 GeV with the exception that the sharp transition (ST) in jet structure near $\nu = 3$ first reported in Ref.~\cite{anomalous} has shifted down to $\nu \approx 2$ at the higher energy, as first noted in Ref.~\cite{tomphenix}. Eq.~(\ref{nchalice}) with $x(\nu)$ defined below was used in this study to relate reported \nch\ values from Ref.~\cite{alicempt} to fractional cross sections and then to Glauber parameters $N_{part}/2$, $N_{bin}$ and $\nu = 2N_{bin}/N_{part}$ according to the methods in Ref.~\cite{powerlaw}.

 %%%%%%%%%%
 %%%%%%%%%%%%
 
 Figure~\ref{alicench} (right panel) shows the $x(\nu)$ trend (dashed curve) that describes the ALICE hadron production data in the left panel (solid curve) defined by
 \bea \label{xnual}
 x(\nu) \hspace{-0.05in}&=&\hspace{-0.05in} 0.028 + 0.141\{1 + \tanh[(\nu - \nu_0)/0.5]\}/2,
 \eea
 where $\nu_0$ = 2 estimates the ST for hadron production in 2.76 TeV \pbpb\ collisions.
The 2.76 TeV $x(\nu)$ expression is divided by factor 1.85 in Fig.~\ref{alicench} for direct comparison with  the 200 GeV trend. We can isolate factor $\langle p_t \rangle_h(\nu)$ by dividing the data in Fig.~\ref{mptaa} (right panel) by $x(\nu)$ from Eq.~(\ref{xnual}). The result (solid points) is described by
\bea \label{mptnu}
\langle p_t \rangle_h(\nu) \hspace{-0.07in}&=&\hspace{-0.07in} 1.00\hspace{-0.02in} +\hspace{-0.02in} 1.70\{1\hspace{-0.02in}-\hspace{-0.02in} \tanh[(\nu - \nu_1)/0.42]  \}/2 
\eea
with $\nu_1 = 1.75$ which defines the solid curve through data. Note that $\langle p_t \rangle_h(\nu)$ in Fig.~\ref{alicench} (right panel) describes an average over $\nu - 1$ secondary \nn\ encounters and for peripheral collisions does not extrapolate to the first-encounter \pp\ value 1.75 GeV/c. The product of Eqs.~(\ref{xnual}) and (\ref{mptnu}) gives the solid curve through data in Fig.~\ref{mptaa} (right panel), and incorporated in Eq.~(\ref{aatcm}) gives the solid curve  through \mpt\ data in the left panel of that figure.

The accurate TCM description of \mpt\ data in Fig.~\ref{mptaa} arises by construction from this analysis procedure. However, the procedure depends on several a priori elements: (a) the \aa\ \mpt\ TCM represented by Eq.~(\ref{aatcm}), (b) a TCM for hadron production represented by Eq.~(\ref{nchalice}) that describes production at any collision energy modulo simple scaling with beam rapidity as discussed in Refs.~\cite{tomphenix,jetmodel} and (c) \pp\ \mpt\ trends accurately represented by a TCM whose energy dependence is consistent with universal jet spectrum properties as described in Ref.~\cite{jetmodel}.

The new information is contained in Eqs.~(\ref{xnual}) and (\ref{mptnu}) which demonstrate a smooth $\tanh(\nu - \nu') $ transition from values in most-peripheral \pbpb\ (essentially \nn) collisions to modified constant values for more-central collisions. The transition in \pbpb\ is centered  near $\nu = 2$ which corresponds to $N_{part}/2 \approx 9$ and $N_{bin} \approx 18$ with fractional cross section $\sigma / \sigma_0 \approx 0.68$. A similar transition is observed in 200 GeV \auau\ near $\nu = 3$ where $N_{part}/2 \approx 30$, $N_{bin} \approx 90$ and $\sigma / \sigma_0 \approx 0.5$. It may be more significant that at the two energies and corresponding path-length values the dijet density is approximately the same, $dn_j / d\eta \approx 2$~\cite{tomphenix}. The hadron production trend at the two energies is related by a simple $\log(\sqrt{s_{NN}})$ energy scaling. The $\langle p_t \rangle_h \approx 1$ GeV/c value for more-central \pbpb\ at 2.76 TeV is the same fraction of the peripheral value (40\%) as that at 200 GeV -- 0.5  vs 1.2 GeV/c.
 
\begin{figure}[h]
  \includegraphics[width=1.65in,height=1.6in]{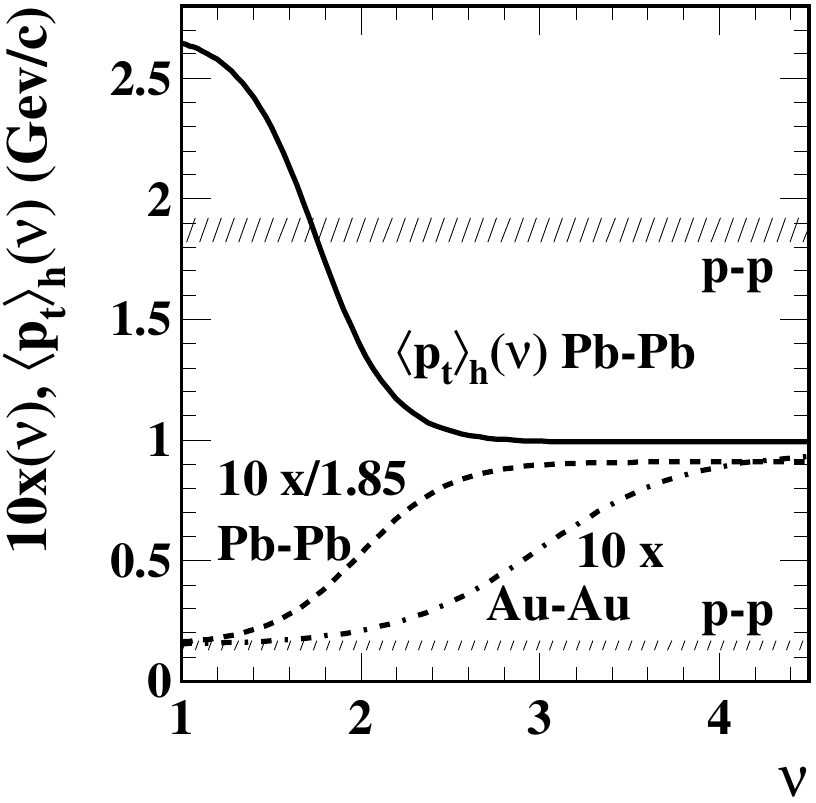}  \includegraphics[width=1.65in,height=1.6in]{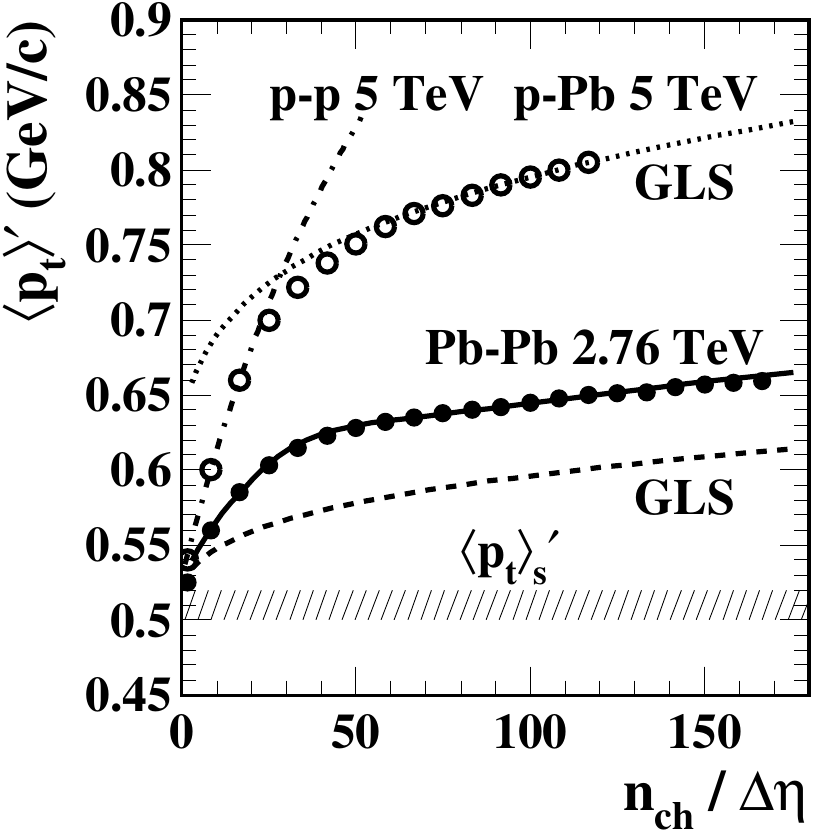}

  \caption{\label{mptppb}
  Left:  The model functions $x(\nu)$ (dashed curve) and $\langle p_t \rangle_h(\nu)$ (solid curve) plotted vs path length $\nu$ over a larger centrality interval ($\nu \approx 6$ for central \aa\ collisions). The  $x(\nu)$ trend for 200 GeV \auau\ collisions (dash-dotted curve) is included for comparison. Corresponding 200 GeV $\langle p_t \rangle_h(\nu)$ data are not currently available.
  Right:   \mpt\ data for 5 TeV \ppb\ collisions from Ref.~\cite{alicempt} (open circles). The dash-dotted curve is the TCM for 5 TeV \pp\ collisions interpolated from the general \pp\ TCM developed in this study. The dotted curve is the GLS for \pbpb\ collisions (dashed curve) with ratio $x$ increased by factor 2.5 and all else  the same.  The \pbpb\ data and TCM from Fig.~\ref{mptaa} (left panel) are included for comparison.
   }  %alice3a, 3e
  \end{figure}

  Figure~\ref{mptppb} (left panel) shows the trends for $\langle p_t \rangle_h(\nu)$ and $x(\nu)$ on path length $\nu$ over a larger centrality interval. The trends for 2.76 TeV \pbpb\ collisions are the solid and dashed curves equivalent to Fig.~\ref{alicench} (right panel). The 2.76 TeV $x(\nu)$ trend is divided by 1.85 for direct comparison with the 200 GeV result. The dash-dotted curve shows $x(\nu)$ for 200 GeV \auau\ collisions with ST near $\nu = 3$ (fractional cross section $\sigma / \sigma_0 \approx 0.5$). There are no equivalent $\langle p_t \rangle_h(\nu)$ data for 200 GeV. We may conclude from these \mpt\ results that aside from translation of the ST from $\nu = 3$ to 2 jet modification in more-central \aa\ collisions at LHC energies is remarkably similar to that at RHIC energies.

  %%%%%%%%%%
  %%%%%%%%%%%%

  %%%%%%%%%%%%%%%
 \section{LHC $\bf p$-$\bf Pb$ data} \label{padat}
   
Figure~\ref{mptppb} (right panel) shows \ppb\ \mpt\ data at 5 TeV~\cite{alicempt} vs \nch (open symbols). Also included for reference are the \pbpb\ results from Fig.~\ref{mptaa} (left panel) and the TCM expression for the \pp\ trend at 5 TeV (dash-dotted curve). The upper GLS curve is the lower GLS curve assuming that constant $x$ increases by factor 2.5 relative to NSD \pp\ but otherwise  jet structure is unchanged in \ppb\ collisions  relative to \pp\ collisions. The \ppb\ collisions are transparent. The \aa\ GLS description of \ppb\ data with eikonal approximation for larger \nch\ is good. The \ppb\ data appear to make a smooth transition from the non-eikonal \pp\ trend to the eikonal \aa\ trend. The transition is located near $n_{ch} / \Delta \eta = 30$.
     
  One can speculate that up to the \ppb\ transition point there is only a single \nn\ collision in peripheral \ppb\  and $\nu \equiv 1$. In  that case the only way to satisfy the increasing \nch\ condition is with increased \nn\ $n_s$ resulting in a large increase in jet production $\propto n_s^2$ due  to the non-eikonal interaction as in single \pp\ collisions.
  At some value of \nch\ the probability of producing a single \nn\ collision with sufficient \nch\ becomes smaller than the probability of a second \nn\ binary collision in more-central \ppb\ collisions and $\nu$ becomes significantly greater than 1. It is then possible to produce more soft hadrons relative to jets by multiple \nn\ collisions, each with a smaller soft multiplicity $n_s$ and therefore jets $\propto n_s^2$. It is interesting that  the transition from non-eikonal to full eikonal behavior apparently occurs within a small \nch\ interval.

%%%%%%%%%%%%%%%
\section{Discussion} \label{disc}

\subsection{Evolution of the HC in \aa\ collisions}

Figure~\ref{mptppb} (left panel) shows the trend of $\langle p_t \rangle_h(\nu)$ vs 2.76 TeV \pbpb\ centrality including a transition from a larger peripheral value to a smaller central value, the ratio being about 2.5. The most rapid variation corresponds to a sharp increase in hadron production or sharp transition near $\nu = 2$. The same phenomenon is observed in hadron yields and spectra for \auau\ at 200 GeV, and a differential analysis of \yt\ spectra for identified hadrons including a two-component decomposition of the hadron spectra~\cite{hardspec} reveals the spectrum hard-component centrality systematics which in turn determine $\langle p_t \rangle_h(\nu)$.

\begin{figure}[h]
  \includegraphics[width=1.65in,height=1.65in]{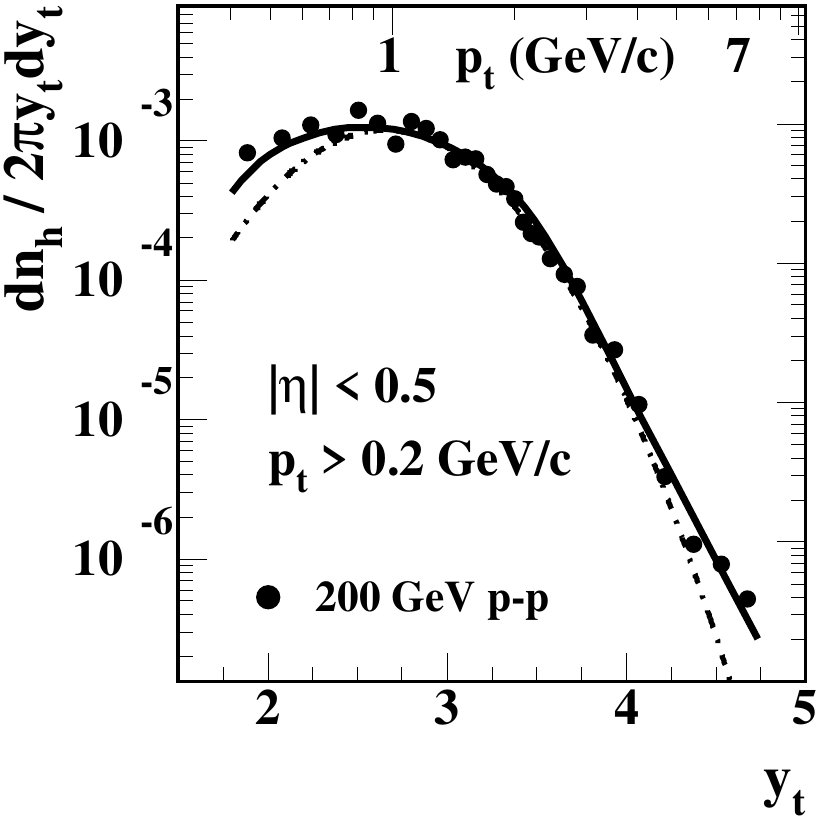}
  \includegraphics[width=1.65in,height=1.65in]{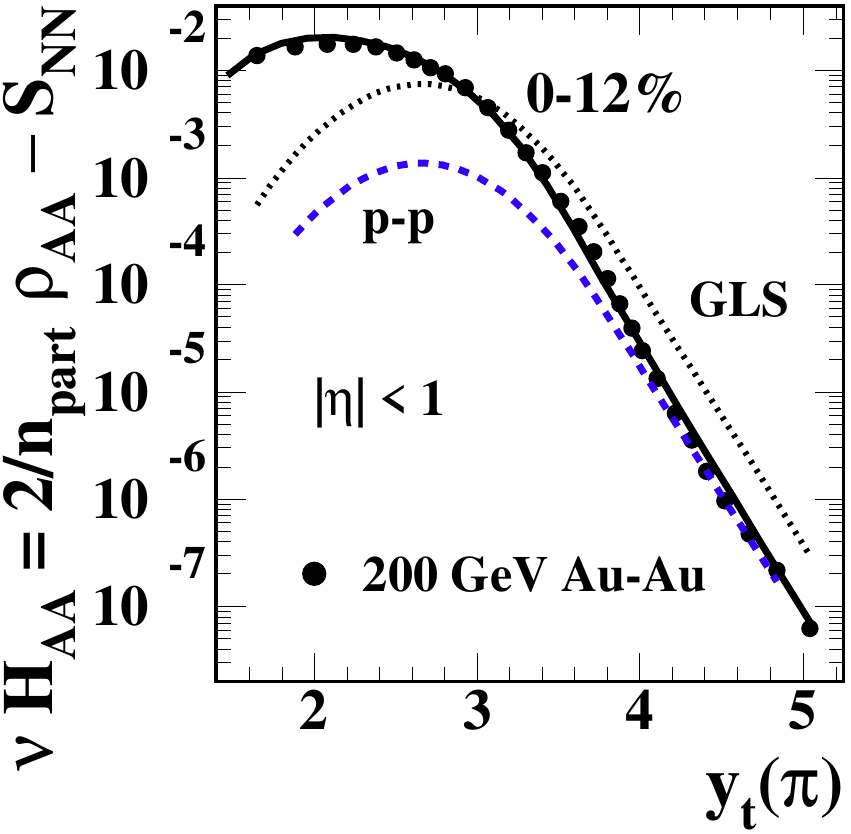}
\caption{\label{specfrag}
Left:  Spectrum hard component from 200 GeV NSD \pp\ collisions (points)~\cite{fragevo}. The solid curve is a pQCD prediction for the corresponding fragment distribution derived from measured fragmentation functions and a dijet total cross section of 2.5 mb.
Right: Spectrum hard component for 0-12\% central 200 GeV \auau\ collisions (points)~\cite{hardspec}. The solid curve is a pQCD prediction based on a simple modification of fragmentation functions~\cite{fragevo}. The dotted curve is a GLS prediction of the TCM extrapolated from \pp\ collisions.
 } % aleph11d2pi, aaspectra10ee
 \end{figure}

Figure~\ref{specfrag} (left panel) shows the non-single-diffractive (NSD) \pp\ spectrum hard component (points) from Ref~\cite{ppprd}. The solid curve is a pQCD calculation of the \pp\ spectrum hard component based on a parton (jet) spectrum bounded below near 3 GeV and integrating to 2.5 mb and measured fragmentation functions~\cite{eeprd,fragevo}. The dash-dotted curve is a Gaussian approximation from Ref.~\cite{ppprd}. The ensemble mean is $\langle p_t \rangle_h \approx 1.2$ GeV/c.

%%%%%%%%%%
%%%%%%%%%%%%

Figure~\ref{specfrag} (right panel) shows the pion spectrum hard component (HC) from 0-12\% central 200 GeV \auau\ collisions (points)~\cite{hardspec}. The dashed curve is the \pp\ Gaussian (with added power-law tail) from the left panel. The dotted curve is a GLS prediction for central \auau\ (assuming \aa\ transparency and $\nu$ increased by factor 5). The solid curve is a pQCD description of the central \auau\ data based on a {\em single modification} of measured fragmentation functions (single-parameter change in a gluon splitting function)~\cite{fragevo}. Fragment reduction at larger $y_t$ is balanced by much larger fragment increase at smaller $y_t$ that conserves the parton energy {\em within resolved jets}~\cite{jetspec}. The large increase in jet fragment production at smaller \yt\ is concealed by the biased ratio measure $R_{AA}$~\cite{nohydro}. The ensemble mean for central collisions is $\langle p_t \rangle_h \approx 0.5$ GeV/c a factor 2.4 smaller than that for \pp\ collisions.
 
The evolution from peripheral HC on the left to central HC on the right proceeds most rapidly about a sharp transition near $\nu = 3$. The ratio of $\langle p_t \rangle_h$ for peripheral vs central is essentially the same as that at 2.76 TeV. In the same  transition interval near $\nu = 3$  200 GeV jet-related angular correlations change rapidly from a Glauber \nn\ linear superposition trend to major changes in certain jet properties~\cite{anomalous}. We thus have strong evidence that the $\langle p_t \rangle_h(\nu)$ trend inferred from 2.76 TeV \pbpb\ collisions reflects parton fragmentation to jets evolving with \aa\ centrality according to pQCD principles in a manner very similar to  that at 200 GeV as in Fig.~\ref{specfrag}.
 
 \subsection{\nn\ first encounters and hadron production} \label{nnfirst}
 
Reference~\cite{kn} describes an oft-cited \aa\ hadron production model based on the assumption that of total \pp\ multiplicity $n_{pp}$ a fraction $x$ arises from ``hard'' processes and a fraction $1-x$ arises from ``soft'' processes, which is the TCM for \pp\ collisions. The result for \aa\ collisions satisfying the eikonal approximation and scaling with the corresponding Glauber parameters is then the expression
 \bea \label{kn1}
 \frac{2}{N_{part}}n_{ch} &=& n_{pp} [1 + x (\nu - 1)]
 \eea
 with $x = n_h / n_{pp}$ which describes hadron production for more-central \auau\ collisions well. However, the implicit assumption that $x$ is the same for all cases is not correct. If $x \approx 0.1$ in more-central \auau\ as observed~\cite{hardspec} it should have the same value for \pp\ collisions. But the observation is $x \approx 0.015$ for 200 GeV \pp\ collisions~\cite{ppprd,jetmodel}.
 
 To resolve the apparent inconsistency we can retreat to the more general expression
 \bea
  \frac{2}{N_{part}}n_{ch} &=& n_s + n_h(\nu) \nu
  \\ \nonumber
  &=& n_s[1 + x(\nu) \nu]
 \eea
 where $x(\nu) = n_h(\nu)/n_s$. Given the assumption that $n_h(\nu)$ is the same for all $\nu$ \nn\ encounters that form does not describe hadron production data for more-central \auau\ collisions unless $x(\nu)$ has a complex structure. In order to arrive at Eq.~(\ref{kn1}) we must assume that in its first \nn\ encounter a nucleon contributes the \pp\ value $n_{h,pp}$ and in $\nu - 1$ subsequent encounters it contributes on average a different value $n_h(\nu)$.  We then have $n_{pp} = n_s + n_{h,pp}$, $x(\nu) =  n_h(\nu)/n_{pp}$ and all aspects are consistent with Eq.~(\ref{kn1}) that describes data.
 
 The success of Eq.~(\ref{kn1}) in describing hadron production data for which $x(\nu)$ varies simply from peripheral (\nn) to central \aa\ collisions implies that the first encounter of a projectile nucleon is special. An encounter with at least one unstruck (unexcited) nucleon produces jet-related hadrons as a \pp\ collision no matter what the A-B environment. In secondary encounters \nn\ collisions produce more jet-related hadrons, the extent depending on the A-B context. The first-encounter effect may reflect the extent to which colored partons are shielded within a nucleon before and after it is excited.
 
\subsection{Comparing theory Monte Carlos with the TCM}

In Fig.~3 of Ref.~\cite{alicempt} several theory Monte Carlo models are compared with the \mpt\ data, and in various ways the MCs fail. We argue that a common problem is the description of jets in high-energy nuclear collisions. In contrast, a simple jet description within the TCM based on pQCD principles and jet measurements provides an accurate and comprehensive description of the \mpt\ data.

The PYTHIA MC includes jet production as a principal mechanism, but based on an eikonal model of colliding composite projectiles that is inconsistent with measured jet production in \pp\ collisions. Thus, in Fig.~3 (top panel) of Ref.~\cite{alicempt} default PYTHIA~\cite{pythia} exhibits the characteristic $n_{ch}^{1/3}$ trend of the eikonal approximation (open diamonds). It is only by adding an ad hoc MPI color reconnection mechanism that the MC results can be made to accommodate the \pp\ \mpt\ data. In contrast, a TCM with  non-eikonal trend derived from \pp\ data describes the \mpt\ data for a broad range of energies in Fig.~\ref{mptdat} (left panel) of the present study to the uncertainty limits of the data. 
An unanticipated result of the TCM \pp\ analysis is the revelation that the energy dependence of $\langle p_t \rangle_h(\sqrt{s})$ is consistent with the energy dependence of minimum-bias jet spectrum widths~\cite{jetmodel}, further buttressing a jet interpretation for the TCM hard component.

Some of the MC models for larger A-B systems are based on PYTHIA and thus inherit the problem of the eikonal approximation for \nn\ collisions. Although that approximation is valid for superposition of \nn\ collisions within \aa\ collisions the \nn\ collisions are not modeled correctly. An example is HIJING~\cite{hijing} based on PYTHIA which fails to describe \auau\ jet-related correlation data within a centrality interval where the data follow a Glauber linear superposition trend~\cite{anomalous}. HIJING correctly implements the Glauber model of \aa\ collisions and could therefore be considered a representative example of the \aa\ TCM, but the PYTHIA modeling of jet production within HIJING \nn\ collisions is incorrect.
The AMPT MC~\cite{ampt} is in turn  based on HIJING but with the addition of final-state parton and hadron rescattering. AMPT inherits the PYTHIA problem within a more-complex parametrized system.

Monte Carlos in which hydrodynamic flow of a thermalized medium plays a dominant role~\cite{epos} cannot describe jet manifestations in more-peripheral \aa\ collisions and therefore cannot provide a comprehensive description of \aa\ collisions. The basis for parametrization of hydro collision models can be questioned if such parametrizations are based on the data system to be described. In contrast, the TCM is required to describe all collision systems consistently with a minimal set of parameters whose values are determined transparently. The present analysis demonstrates that a TCM based on the universality of  jets leads to comprehensive and accurate descriptions of a broad range of nuclear collision data and new insights into jet formation.

%   Generally, describe Pt, nch and Pt/nch = \mpt\ on ns, nu and root-s. Simply and consistently for \pp\ and \aa\
%    Plot Pt vs nu and nch vs nu separately, can hydro make that Pt?

\subsection{What can be learned from the LHC $\bf \langle p_t \rangle$ data?}

Viewed within a TCM context the recent LHC \mpt\ data are quite informative. A TCM based on simple jet contributions  is found to describe \mpt\ data accurately and consistently for a variety of collision systems. Alternative MC models strongly disagree with the data. Effective spectrum \pt\ cutoffs can be inferred accurately from \mpt\ vs \nch\ trends combined with the TCM spectrum soft-component model. The TCM soft component is universal, remains the same for all A-B combinations over a large energy interval. The non-eikonal nature of jet production in \pp\ collisions is confirmed by \mpt\ data.

The \pp\ \mpt\ hard component $\langle p_t \rangle_h(\sqrt{s})$ is observed to scale linearly with $\Delta y_{max} = \ln(\sqrt{s} / \text{3 GeV})$  representing the widths of MB jet spectra from 200 GeV to 7 TeV~\cite{jetmodel}. The linear relation is just that expected if the TCM spectrum hard component represents jet fragments as described by pQCD~\cite{fragevo}. That result is in turn consistent with the \mpt\ hard component amplitude scaling as $n_h \propto (\Delta y_b)^2 = [\ln(\sqrt{s} / \text{10 GeV})]^2$ implying non-eikonal jet production in \pp\ collisions~\cite{jetmodel}.

The $n_h(\nu)$ and $\langle p_t \rangle_h(\nu)$ centrality trends in 2.76 TeV \pbpb\ are generally consistent with results from 200 GeV \auau\ collisions. Spectrum evolution in the latter system follows a pQCD description of FFs in which FF evolution tends to conserve the leading-parton energy but with a shift to lower-momentum fragments. A single QCD parameter describes the FF evolution in \aa.

From the variation with centrality of hadron production in 200 GeV \auau\ and 2.76 TeV \pbpb\ collisions we already see evidence for a difference between the first \nn\ encounter and subsequent \nn\ collisions. Applying the same approach to total hard $P_{t,h}$ production reveals a similar trend in 2.76 TeV \pbpb\ collisions. In peripheral \aa\ collisions the second \nn\ encounter produces more $P_{t,h}$, and thereafter the $P_{t,h}$ per \nn\ collision increases to a larger constant value. The ratio of peripheral to central \aa\ $\langle p_t \rangle_h$ values is similar ($\approx 2.5$) at the two energies.
  
Finally, in 5 TeV \ppb\ collisions we observe a smooth transition from \pp\ non-eikonal to \aa\ eikonal  jet-related trends with increasing \nch, possibly a simple matter of competing probabilities for two production mechanisms at a given total \nch.

When combined with previous measurements of yields spectra and correlations in a TCM context the LHC \mpt\ data provide strong evidence that in any collision system jet production is responsible for nearly all $P_t$ and hadron production exceeding a universal soft component.

%%%%%%%%%%%
\section{Summary}\label{summ}

Recent measurements of event-ensemble mean transverse momentum $\langle p_t \rangle$ vs charged-hadron multiplicity $n_{ch}$ for $p_t$ spectra from 5 TeV p-Pb and 2.76 TeV Pb-Pb collisions and from p-p collisions for several energies were compared to several theory Monte Carlos. In most cases there are severe disagreements between theory and data, leaving interpretation of the \mpt\ data unresolved.

In the present study we develop a two-component model (TCM) for total $P_t$ production (within some angular acceptance) in high energy nuclear collisions based on previous measurements of dijet production, parton fragmentation and jet spectra. The underlying assumption of the TCM is that of two contributions to $P_t = P_{t,s} + P_{t,h}$ the soft component is a universal feature of high energy collisions corresponding to longitudinal dissociation of participant nucleons. The complementary hard component is entirely due to minimum-bias jet production in \nn\ collisions. The same model is applied to charged-hadron production in the form $n_{ch} = n_s + n_h$. 

Given TCMs for $P_t$ and $n_{ch}$ we obtain the TCM for $\langle p_t \rangle$ as $P_t / n_{ch}$. A key feature of the \mpt\ TCM is the observation that jet production in \pp\ collisions does not follow a semiclassical eikonal approximation as commonly assumed in \pp\ Monte Carlo models. The observed non-eikonal production trend is incorporated into the TCM for this study. Models for \pp\ and \aa\ collisions are derived separately depending on the relation to the eikonal approximation and applicability of the Glauber model to \aa\ collisions. To aid data comparison we develop a correction for biases due to incomplete \pt\ acceptance.

Aside from confirming the non-eikonal nature of \pp\ jet production the main result of the \pp\ TCM analysis is extraction of a \mpt\ hard component $\langle p_t \rangle_h(\sqrt{s})$ for several collision energies. The energy variation of the hard component is found to be linearly related to parameter $\Delta y_{max} = \ln(\sqrt{s} / \text{3 GeV})$ which, in a separate study of jet systematics, is observed to describe the energy dependence of minimum-bias jet spectrum widths. The linear relation is understandable in a context where $\langle p_t \rangle_h$ is determined by a hadron spectrum hard component composed of jet fragments described by folding a fragmentation-function ensemble with a minimum-bias jet spectrum. The jet interpretation is strongly supported.

The \pbpb\ \mpt\ analysis has several implications. The $P_{t,h}$ production generally follows binary-collision scaling according to the eikonal approximation as expected for jet production in \aa\ collisions. However, as for hadron production the data suggest that the first \nn\ encounter of a nucleon participant in an \aa\ collision is special, yielding the same results as a \pp\ collision. Subsequent encounters produce different results (more hadron fragments per \nn\ encounter, reduced $\langle p_t \rangle_h$) with increasing \aa\ centrality. The increase of $n_h$ and decrease of $\langle p_t \rangle_h$ is consistent with modifications to jet fragment distributions (hard components) observed in hadron \pt\ spectra. The systematics of jet modification in 2.76 TeV \pbpb\ collisions appears to be very similar to those in 200 GeV \auau\ collisions, the only differences being (a) the location of a sharp transition from \nn\ linear superposition (\aa\ transparency) to strongly modified jets and (b) increase of soft multiplicity and hard jet-related components by factors $\ln(\sqrt{s_{NN}}/\text{10 GeV})$ and $\ln^2(\sqrt{s_{NN}}/\text{10 GeV})$ respectively as extrapolated from spectrum and correlation systematics at RHIC energies.

Generally speaking the TCM for \mpt\ data provides a very compact data representation. Much of the model is determined by prior measurements of other aspects of high energy nuclear collisions. The novelty arising in the \mpt\ study is (a) the energy dependence of $\langle p_t \rangle_h(\sqrt{s})$ for \pp\ collisions found to comply with previously measured jet spectrum systematics, (b) the participant first-encounter effect already implicit in hadron production systematics and (c) variation of $\langle p_t \rangle_h(\nu)$ with \pbpb\ centrality which is qualitatively consistent with jet modifications in 200 GeV \auau\ collisions described by pQCD.

The \ppb\ result which is said to present conceptual difficulties appears to be consistent with the TCMs for \pp\ and \aa\ collisions, transitioning smoothly from the \pp\ trend at lower multiplicities (quantitative agreement) to a Glauber linear superposition model of \aa\ collisions with no modification of jet fragmentation in more-central collisions (close agreement with the functional form).

%%%%%%%%%%%%%%%%%%
This work was supported in part by the Office of Science of the U.S.\ DOE under grant DE-FG03-97ER41020.

%%%%%%%%%%%%%%%%%%%%%%%%%%%%

\end{document}